\documentclass[aps,prl,superscriptaddress,twocolumn]{revtex4}
\usepackage{graphicx}
\usepackage{amsmath}
\usepackage{amssymb}
\usepackage{booktabs}	

\begin{document}

\title{Photodissociation of a diatomic molecule in the quantum regime reveals \\ ultracold chemistry}

\author{M. McDonald}
\affiliation{Department of Physics, Columbia University, 538 West 120th Street, New York, NY 10027-5255, USA}
\author{B. H. McGuyer}
\affiliation{Department of Physics, Columbia University, 538 West 120th Street, New York, NY 10027-5255, USA}
\author{F. Apfelbeck}
\altaffiliation[Present address:  ]{Faculty of Physics, Ludwig Maximilian University of Munich, Schellingstrasse 4, 80799 Munich, Germany}
\affiliation{Department of Physics, Columbia University, 538 West 120th Street, New York, NY 10027-5255, USA}
\author{C.-H. Lee}
\affiliation{Department of Physics, Columbia University, 538 West 120th Street, New York, NY 10027-5255, USA}
\author{I. Majewska}
\affiliation{Quantum Chemistry Laboratory, Department of Chemistry, University of Warsaw, Pasteura 1, 02-093 Warsaw, Poland}
\author{R. Moszynski}
\affiliation{Quantum Chemistry Laboratory, Department of Chemistry, University of Warsaw, Pasteura 1, 02-093 Warsaw, Poland}
\author{T. Zelevinsky}
\email{tz@phys.columbia.edu}
\affiliation{Department of Physics, Columbia University, 538 West 120th Street, New York, NY 10027-5255, USA}

\begin{abstract}     
Chemical reactions at temperatures near absolute zero require a full quantum description of the reaction pathways and enable enhanced control of the products via quantum state selection.  Ultracold molecule experiments have provided initial insight into the quantum nature of basic chemical processes involving diatomic molecules, for example from studies of bimolecular reactions \cite{YeOspelkausScience10_KRbReactions}, but complete control over the reactants and products has remained elusive. The ``half-collision" process of photodissociation is an indispensable tool in molecular physics and offers significantly more control than the reverse process of photoassociation \cite{JonesRMP06}.  Here we reach a fully quantum regime with photodissociation of ultracold $^{88}$Sr$_2$ molecules where the initial bound state of the molecule and the target continuum state of the fragments are strictly controlled.  Detection of the photodissociation products via optical absorption imaging reveals the hallmarks of ultracold chemistry:  resonant and nonresonant barrier tunneling, importance of quantum statistics, presence of forbidden reaction pathways, and matter wave interference of reaction products.  In particular, this interference yields fragment angular distributions with a strong breaking of cylindrical symmetry, so far unobserved in photodissociation.  We definitively show that the quasiclassical description of photodissociation fails in the ultracold regime.  Instead, a quantum model accurately reproduces our results and yields new intuition.
\end{abstract}
\date{\today}
\maketitle

Photodissociation is a chemical reaction where one or several photons split a molecule into fragments with relative velocities that conserve energy and momentum.  It is an important process in nature, affecting the composition of interstellar clouds and enabling biological photosynthesis.  In the laboratory, it is a powerful method for studying molecular bonding, since the character of the bond-breaking transition can be deduced from both the fragment velocities and their angular distribution.  Angular distributions are rich observables and are important, for example, in photoionization experiments \cite{ReidARPC03_PhotoelectronAngularDistributions} where they provide a route to ``complete" measurements of the ionization matrix element amplitudes and phases \cite{HockettPRL14_CompletePhotoionizationExperiments}.  It was realized over 50 years ago that for diatomic molecules the fragment angular distribution produced by single photon dissociation could be described by $I(\theta)=I_0[1+\beta_{20}P_2^0(\cos\theta)]$, where $\theta$ is the polar angle relative to the quantum axis, $P_l^m(\cos\theta)$ is an associated Legendre polynomial, and $\beta_{20}\in[-1,2]$ parametrizes the degree of anisotropy \cite{HerschbachZarePIEEE63_DiatomicPhotodissociation,ZareMPC72_PhotoejectionDynamics}.  This description has worked well for molecules prepared in spherically symmetric states.  However, if a molecule can be prepared in an arbitrary quantum state, for example, with angular momentum $J_i$ and its projection $M_i$, then quantum mechanics allows for more complex distributions.  In most experiments, these distributions agree with a quasiclassical description replacing $I_0$ with $I_0(\theta,\phi)=|\Phi_i(\theta,\phi)|^2$, where $|\Phi_i|^2$ is an angular probability density for the initial molecular axis orientation and $\phi$ is the azimuthal angle \cite{BernsteinChoiJCP86_StateSelectedPhotofragmentation,ZareCPL89_PhotofragmentAngularDistributions,Supplemental}.
This description of fragmentation is intuitive because it involves the product of a correction for a non-spherical prepared molecule with a probability density of photon absorption, but its applicability to fragmentation in the quantum regime has been questioned over the years \cite{SeidemanCPL96_MagneticStateSelectedPDDistributions}.

Our experiments demonstrate that for fully quantum photodissociation the fragment angular distributions are determined entirely by the final (continuum) states, and are generally inconsistent with the quasiclassical description.  For each electronic channel, the measured distribution is an intensity (or differential cross section),
\begin{equation}
I(\theta,\phi)=|f(\theta,\phi)|^2,
\label{eq:IPsiRot}
\end{equation}
that is the square of a scattering amplitude $f$ that can be expanded in terms of partial amplitudes, $f(\theta,\phi)=\sum_{JM}f_{JM}\psi_{JM}(\theta,\phi)$, using angular basis functions $\psi_{JM}$ of the outgoing channel.  The intensities for individual electronic channels superpose to produce the total $I(\theta,\phi)$ \cite{Supplemental}, and the amplitudes $f_{JM}$ are obtained by connecting the bound state molecular wave function to the continuum state wave function via Fermi's golden rule \cite{Supplemental}.

Cylindrically asymmetric distributions with $\phi$ dependence are possible if several $M$ states are coherently created, since $\psi_{JM}(\theta,\phi)\equiv e^{iM\phi}\psi_{JM}(\theta,0)$ \cite{Supplemental}.  The angular distributions can be summarized with the parametrization
\begin{equation}
I(\theta,\phi)\propto1+\sum_{l=1}^{\infty}\sum_{m=0}^{l}\beta_{lm}\cos(m\phi)P_l^m(\cos\theta),
\label{eq:IBeta}
\end{equation}
where for homonuclear diatomic molecules $l$ is restricted to even values.
The $\beta$ coefficients are directly related to the amplitudes $f_{JM}$ \cite{Supplemental}, but hide some of the symmetries that are apparent from using the representation of Eq. (\ref{eq:IPsiRot}).

\begin{figure}
\includegraphics*[trim = 0in 0in 0in 0in, clip, width=3.5in]{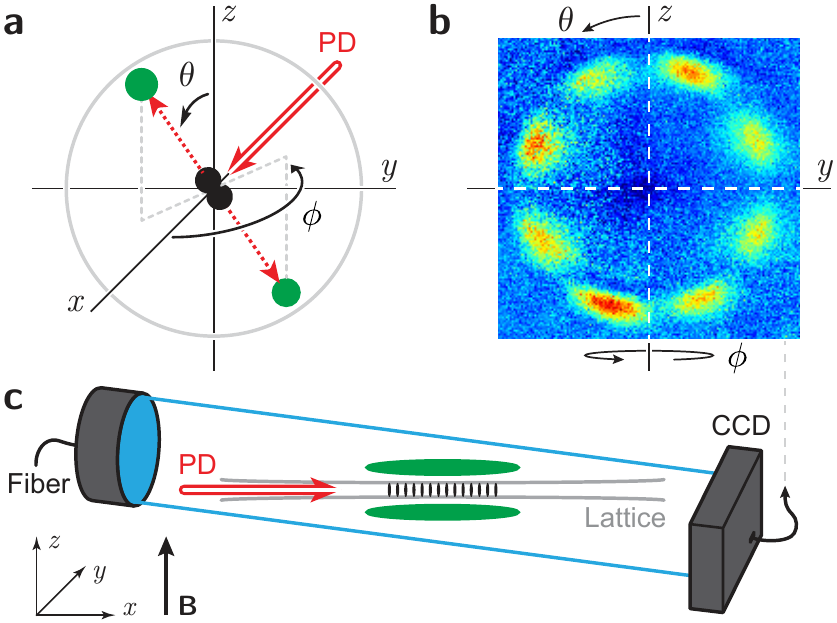}
\caption{Photodissociation of diatomic molecules in an optical lattice.  a) A Newton sphere is formed by the fragments of a dissociating homonuclear molecule with a well controlled fragment speed.  The position of a fragment on the sphere is parametrized by the polar angle $\theta$ relative to the $z$ axis, and by the azimuthal angle $\phi$ relative to the $x$ axis in the $xy$ plane.  The photodissociating laser propagates along $+x$.  b) A fragment pattern obtained via absorption imaging corresponds to the Newton sphere projected onto the $yz$ plane.  This pattern distinctly shows a $\phi$ dependence in addition to $\theta$ dependence, i.e. a lack of cylindrical symmetry about the $z$ axis.  c) The fragment angular patterns are obtained with a CCD camera positioned coaxially with the 1D optical lattice.  The imaging beam with a large waist size is delivered via optical fiber and detects the Sr atomic fragments by absorption.  The photodissociating (PD) laser is coaligned with the lattice axis $x$, and the imaging laser is nearly coaligned with $x$ (a small tilt is present for technical reasons).  A magnetic field can be applied along the $z$ axis.}
\label{fig:Fig1}
\end{figure}
To attain full control over the initial and final quantum states, we trap diatomic strontium molecules, $^{88}$Sr$_2$, in a 1D optical lattice at a temperature of $5$ $\mu$K \cite{ZelevinskyMcGuyerNPhys15_Sr2M1}.  The lattice wavelength is $910$ nm and the trap depth is $50$ $\mu$K.  We photodissociate the molecules with linearly polarized narrow-bandwidth light that propagates along the lattice axis, and detect the fragments via absorption imaging to produce a 2D projection of the 3D spherical shell (``Newton sphere") formed by the fragments.  The duration of the photodissociation light pulse is typically $20$ $\mu$s, to optimize both the signal to noise ratio and the angular resolution.  A complementary spectroscopic detection method counts the total number of atomic fragments produced immediately after the application of the photodissociation pulse.  A small bias magnetic field $B$ sets the quantum axis and controls Zeeman shifts.  We probe continuum energies in the range of $0$-$15$ mK because it matches the typical electronic and rotational barrier heights.  We have confirmed that our results are unaffected by the small lattice trap depth.  The experimental geometry is illustrated in Fig. \ref{fig:Fig1}.

\begin{figure}
\includegraphics*[trim = 0in 0in 0in 0in, clip, width=3.5in]{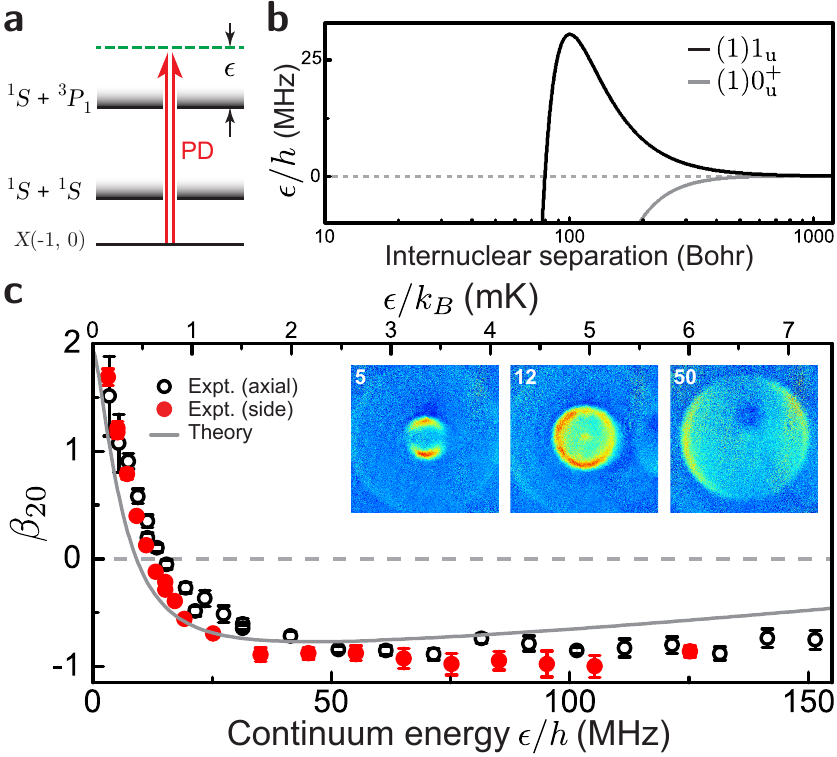}
\caption{Photodissociation to a multichannel continuum.  a) Schematic diagram showing the continuum energy $\epsilon$ and the initial molecular state X$(v_i,J_i)$.  b) Potential energy structure $\lesssim1$ mK of the $^1S+{^3P}_1$ continuum, showing both electronic potentials ($0_u^+$ and $1_u$) that are coupled to the ground state via E1 transitions \cite{KillianBorkowskiPRA14_SrPAMassScaling}.  c) The angular anisotropy parameter $\beta_{20}$ for $^{88}$Sr$_2$ photodissociation to the $^1S+{^3P}_1$ continuum, including two types of experimental methods (using axial-view and side-view CCD camera positioning) and theory based on a quantum chemistry model.  The insets show photofragment images at three different continuum energies.  The images and the curves indicate a steep change of angular anisotropy in the 0-2 mK continuum energy range.  The experimental error bars for axial imaging arise from applying the pBasex algorithm with several choices of center points and averaging the results, and for side-view imaging from least squares fits to Eq. (\ref{eq:IBeta}) convolved with a blurring function accounting for experimental imperfections.  For this measurement, the applied field is $B=0$.}
\label{fig:Fig2}
\end{figure}
Anisotropic angular distributions of molecular fragments were first observed nearly 50 years ago \cite{SolomonJCP67_PhotodissociationPhotolysisMapping}.  Since then, photodissociation has emerged as a workhorse of physical chemistry, but the lack of sufficiently cold molecular samples has precluded its wide use for studies of quantum behavior near dissociation thresholds.  To investigate a multichannel electronic continuum, we prepared the ultracold molecules in the $J_i=0,\;M_i=0$ state of the weakly bound vibrational level $v_i=-1$ of the ground state potential X (below the $^1S+{^1S}$ threshold), and coupled them to the excited $^1S+{^3P}_1$ continuum with 689 nm photodissociation light (negative vibrational level indices count down from the dissociation limit).  The subsequent molecular fragmentation is the electric dipole (E1) process illustrated in Fig. \ref{fig:Fig2}(a).  If the photodissociation light polarization is parallel to the quantum axis ($p=0$), the fragments can only have $J=1,\;M=0$, because $J\geq1$ for the $0_u^+$ and $1_u$ electronic potentials shown in Fig. \ref{fig:Fig2}(b).  Since $1_u$ has a $\sim30$ MHz ($\sim1.5$ mK) repulsive electronic barrier, we expect the fragment angular distribution to evolve in the probed energy range due to barrier tunneling.  Indeed we observe a steep variation of the single anisotropy parameter needed to describe this particular process, $\beta_{20}$ from Eq. (\ref{eq:IBeta}).  Two methods were used to acquire and process this fragmentation data:  axial-view imaging and processing with the pBasex algorithm \cite{GarciaRSI04_2DImageInversion}, and side-view imaging followed by curve fitting to an integrated density \cite{Supplemental}.  Figure \ref{fig:Fig2}(c) shows agreement of the axial-view and side-view imaging approaches and reveals the fragment pattern evolution from a parallel dipole ($\beta_{20}\sim2$, $\epsilon/h\sim5$ MHz), to a sphere ($\beta_{20}\sim0$, $\epsilon/h\sim12$ MHz), and to a perpendicular dipole ($\beta_{20}\sim-1$, $\epsilon/h\sim50$ MHz).  A quantum chemistry model of Sr$_2$ \cite{MoszynskiSkomorowskiJCP12_Sr2Dynamics,KillianBorkowskiPRA14_SrPAMassScaling} was used to calculate the expected anisotropy \cite{Supplemental}, as also plotted in Fig. \ref{fig:Fig2}(c).  The theory shows good qualitative agreement with experiment.  The theoretical $0_u^+$ and $1_u$ Coriolis-mixed potentials agree well with high-precision bound-state $^{88}$Sr$_2$ spectroscopy \cite{ZelevinskyMcGuyerPRL13_Sr2ZeemanNonadiabatic,ZelevinskyMcGuyerNPhys15_Sr2M1,ZelevinskyMcGuyerNJP15_Sr2Spectroscopy}, but this work is the first rigorous test of their predictive power above dissociation thresholds.

\begin{figure}
\includegraphics*[trim = 0in 0in 0in 0in, clip, width=3.5in]{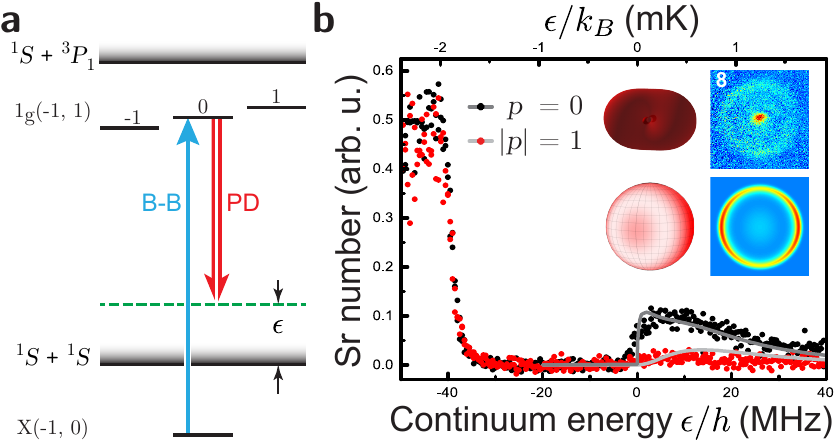}
\caption{Electric-dipole forbidden photodissociation experiment and theory.  a) Molecules in $M_i=0$ of a long-lived gerade state below the $^1S+{^3P}_1$ threshold are prepared with a bound-bound (B-B) laser and fragmented at the gerade ground state continuum with a photodissociation (PD) laser.  b) The line shape of photodissociation that proceeds via the higher-order M1 and E2 couplings is on the right side of the spectrum.  The left portion is strong E1 dissociation to the $^3P_1+{^3P}_1$ continuum.  As predicted, the forbidden photofragmentation line shape is strongest for $p=0$, where in this case the magnetic field of the PD light is parallel to the quantum axis.
Four images are shown as insets.  The upper left maps the predicted fragment detection probability at each angle to the radial coordinate of a surface, while the lower left maps it to a brightness on the Newton sphere (as if the fragments were coating the sphere's inner surface).  At lower right is the 2D projection of the simulated Newton sphere incorporating experimental blur, and at upper right is the directly comparable experimental image (in this case, 8 MHz above threshold).  The small bright dot at the center of the upper right inset results from spontaneous photodissociation of the metastable molecules to low-energy atoms that are recaptured by the lattice.  This decay slightly reduced the resolution of the images.}
\label{fig:Fig3}
\end{figure}
Electric-dipole forbidden photodissociation processes are important in atmospheric physics as they relate to the molecular oxygen Herzberg continuum problem to which extensive effort has been directed \cite{GibsonBuijsseJCP98_AngularDistrHerzbergContinuum}.  However, forbidden photodissociation based purely on magnetic dipole (M1) and electric quadrupole (E2) processes has not been previously observed.  In most cases, the E1 process is present in addition to M1/E2, making it challenging to study the very weak processes independently.  Our ultracold Sr$_2$ setup allows sensitive measurements of pure M1/E2 fragmentation, either with high resolution spectroscopy or via absorption imaging of fragment angular distributions, accompanied by quantum mechanical calculations.  Using resonant $\pi$ pulses, we prepare the $M_i=0$ component of the subradiant $1_g(v_i=-1,J_i=1)$ metastable molecular state that has no E1 coupling to the ground state \cite{ZelevinskyMcGuyerNPhys15_Sr2M1}, as shown in Fig. \ref{fig:Fig3}(a).  The frequency of a linearly polarized dissociation laser was varied as shown in Fig. \ref{fig:Fig3}(b).  The prominent feature on the left side of the spectrum results from E1 fragmentation to the higher-lying $^3P_1+{^3P}_1$ continuum and is present for both polarization angles, while the weaker, polarization-dependent feature on the right side is the M1/E2 fragmentation line shape.  The insets show the predicted and measured angular distribution near the peak of the spectrum, which is qualitatively different from all E1 cases observed in this work.
This is a consequence of $M$ selection rules for E2 transitions that create final states with $|\Delta M|=4$ (as opposed to our E1 experiments where the maximum $|\Delta M|=2$).  The data is in reasonably good agreement with predictions for both laser polarizations.
As the spectrum shows, this forbidden process tapers off rapidly and has a substantial cross section only for submillikelvin product energies.
While the $p=0$ photodissociation spectrum is unaffected by the M1/E2 transition moment interference (selection rules ensure access to different $M$ components by the M1 and E2 processes), the fragment angular distributions are sensitive to this energy-dependent interference \cite{Supplemental}.

\begin{figure}
\includegraphics*[trim = 0in 0in 0in 0in, clip, width=3.5in]{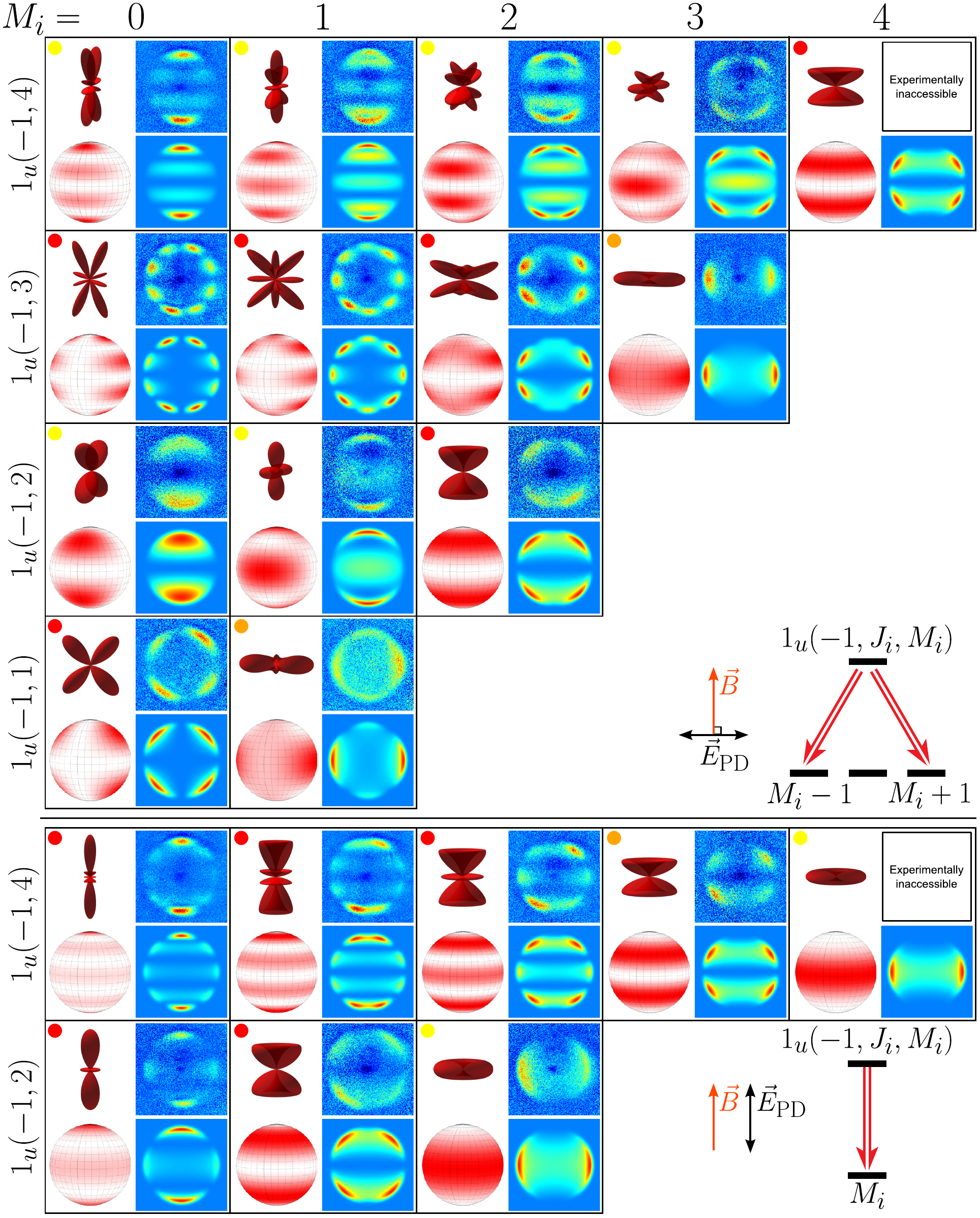}
\caption{Absorption images and simulations of weakly bound $^{88}$Sr$_2$ near the $^1S+{^3P}_1$ threshold fragmenting to the ground state at continuum energies of several mK.  The upper and lower sets correspond to PD laser polarizations $|p|=1$ and $0$, respectively, and the initial states are $1_u(v_i,J_i)$ as indicated.  (Photodissociation of $0_u^+$ molecules has been similarly studied \cite{Supplemental}.)  For each $(J_i,M_i)$, four images are shown, as in Fig. \ref{fig:Fig3}(b).
The $M_i=4$ states were not possible to populate experimentally.  For $|p|=1$, matter wave interference occurs when two $M$ components emerge, leading to direct observation of strongly $\phi$ dependent patterns.  For each case, the degree of agreement with the quasiclassical approximation is indicated by a colored dot, as explained in the text.}
\label{fig:Fig4}
\end{figure}
We take advantage of the single-channel ground state potential of the spinless $^{88}$Sr$_2$ to test validity of the quasiclassical description of photodissociation, to explore chemistry in the ultracold limit, and to obtain a library of fragment angular distributions corresponding to a full set of final quantum states.  We optically populate individual ($1\leq J_i\leq4$), $M_i$ molecular states below the $^1S+{^3P}_1$ threshold and immediately fragment them at the $^1S+{^1S}$ ground state continuum with linearly polarized light, in several cases applying $B$ up to $20$ G to enable symmetry forbidden transitions \cite{ZelevinskyMcGuyerPRL15_Sr2ForbiddenE1}.  Quantum statistics of identical bosons ensure that only even $J$ values are allowed for $^{88}$Sr$_2$, permitting us to utilize E1 selection rules to obtain final states with unique $J$ (e.g. starting from $J_i=2$ or $4$).  Furthermore, if multiple ``partial waves" with different $J$ interfere, typically a single $J$ wave strongly dominates at certain continuum energies, as discussed below.  Selection rules ensure that $M=M_i$ for $p=0$ and $M=M_i\pm1$ for $|p|=1$, where $|p|=1$ denotes linear polarization of the photodissociation laser that is perpendicular to the quantum axis.  Thus, the $J$ selectivity allows us to image pure $M$ state rotational probability densities, as well as their quantum interferences.  Figure \ref{fig:Fig4} shows a full range of possible fragment angular distributions given by Eq. (\ref{eq:IPsiRot}) if $f(\theta,\phi)=Y_{J,M_i}(\theta,\phi)$ or $\sqrt{R}\,Y_{J,M_i-1}(\theta,\phi)+e^{i\delta}\sqrt{1-R}\,Y_{J,M_i+1}(\theta,\phi)$.  Here the coherent superposition is possible only if $|p|=1$, the independent amplitude and phase parameters are $R$ and $\delta$, and the spherical harmonic $Y_{JM}=\psi_{JM}$ for the ground continuum.  (At the continuum energies chosen here, the $p=0$ patterns for $J_i=1,3$ would be nearly redundant with $J_i=2,4$ and are thus omitted.)  We emphasize that the clean observations of the cylindrical-symmetry-breaking coherences between the outgoing spherical harmonics are possible because we can controllably create coherent pairs of $M$ states.

Figure \ref{fig:Fig4} suggests the following observations:  (i) If selection rules allow only a single value of $M$, the fragment angular distributions are cylindrically symmetric, and $\phi$ dependence is not possible.  Conversely, if two $M$ values participate in photodissociation, the fragment probability densities interfere in a way that breaks the cylindrical symmetry and produces $\phi$ dependent patterns.  Figure \ref{fig:Fig4} illustrates multiple cases of a diatomic molecule fragmenting into up to eight directions defined by distinct $(\theta,\phi)$ regions.  (ii)  The breakdown of the quasiclassical description of photodissociation is apparent, as indicated by the colored dots.  A yellow dot indicates qualitative agreement with the quasiclassical approximation (that cannot be made exact by adjusting $\beta_{20}$ from its axial recoil values of $2$ or $-1$ \cite{Supplemental}), while an orange dot indicates disagreement that can become a qualitative agreement by adjusting $\beta_{20}$.  A red dot indicates clear disagreement for all $\beta_{20}$, usually because fragments are observed where $|\Phi_i|^2$ has a node.
All the $1_u$ cases fail the quasiclassical interpretation to varying degrees.  While this could be expected \cite{ZareBeswickJCP08_PhotofragmentAngularDistrQuantClass}, surprisingly even photodissociation of $0_u^+$ molecules fails the quasiclassical model in all cases where more than a single $J$ is present in the final state \cite{Supplemental}.  (iii) For $|p|=1$, $M_i=0$, and $J_i=4,\;3$, the same final states ($J=4,\;M=\pm1$) are produced for the continuum energies used in Fig. \ref{fig:Fig4}.  Thus, according to Eq. (\ref{eq:IPsiRot}), we could expect to observe identical fragment patterns.  However, a subtle point is that one-photon coupling from an odd $J_i$ produces $M=M_i\pm1$ probability amplitudes with an opposite relative phase than from an even $J_i$.  This results in identical $\phi$ dependent patterns that are rotated by $90^{\circ}$ relative to each other.  The same mapping of the relative phase onto the rotation angle of the matter wave interference pattern occurs for $|p|=1$, $M_i=0$, and $J_i=2,\;1$.  (iv) The previous point roughly holds for the higher values of $M_i$ as well, but slightly different populations of $M=M_i\pm1$ are produced due to asymmetrical coupling strengths.  For example, the interference patterns for $(J_i,M_i)=(4,2)$ and $(3,2)$ are not only rotated relative to each other, but have slightly different shapes.

\begin{figure}
\includegraphics*[trim = 0in 0in 0in 0in, clip, width=3.5in]{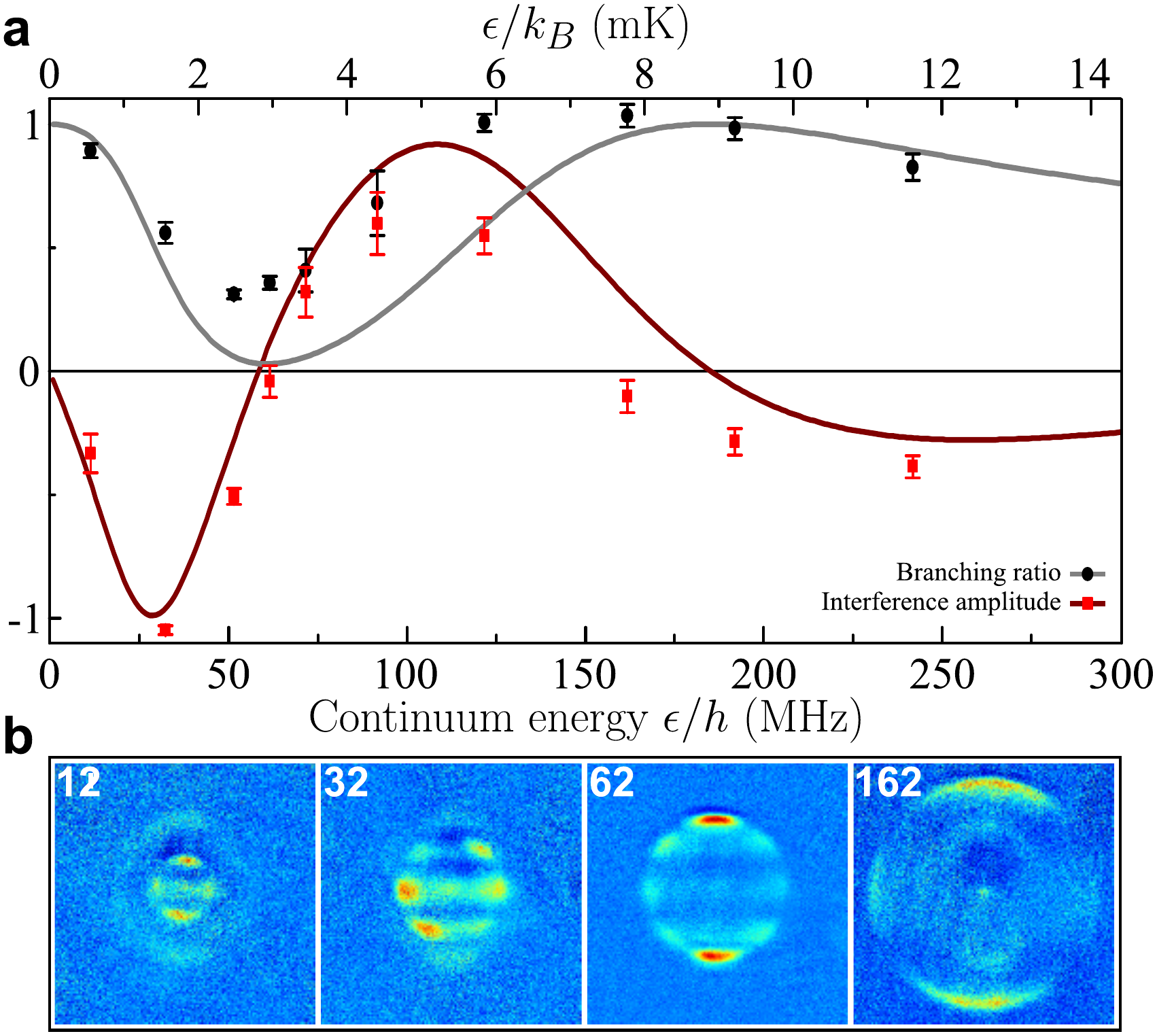}
\caption{Energy dependence of fragment angular distributions.  The $0_u^+(v_i=-3,J_i=3,M_i=0)$ molecules are photodissociated to the ground state continuum.  Optical selection rules ensure that $M=0$ and $J=2,4$.  a) Energy evolution of the branching ratio $R$ and of the interference amplitude between $J=2,4$.  The data is analyzed with pBasex and error bars are estimated by taking the spread of the values resulting from the uncertainty of the effective saturation intensity when processing absorption images.  The theoretical predictions are calculated with the quantum chemistry model.  The plot reveals a $J=4$ (``$g$-wave") shape resonance at $\sim3$ mK.  b) Fragment images at various continuum energies, showing the evolution of the angular pattern.  The faint anisotropic, energy independent pattern with roughly the same radius as in the 62 MHz image arises from spontaneous photodissociation into the shape resonance.}
\label{fig:Fig5}
\end{figure}
Ultracold photodissociation readily reveals features of the continuum just above threshold.  The ability to freely explore a large range of continuum energies, coupled with strict optical selection rules and the preparation of single quantum states, provides a versatile tool to isolate and study individual reaction channels.  While Fig. \ref{fig:Fig2} explored tunneling through an electronic barrier, Fig. \ref{fig:Fig5} shows the evolution of the fragment angular distributions when only rotational barriers are present.  Here, $^{88}$Sr$_2$ molecules in the $0_u(v=-3,J_i=3,M_i=0)$ state are dissociated with $p=0$, resulting in continuum states with $M=0$ and $J=2,4$.  This mixture can be described with two independent parameters as $f(\theta,\phi)=\sqrt{R}\,Y_{20}(\theta,\phi)+e^{i\delta}\sqrt{1-R}\,Y_{40}(\theta,\phi)$.  Figure \ref{fig:Fig5}(a) shows the plot of the branching ratio $R$, as well as of the interference amplitude $2\cos\delta\sqrt{R(1-R)}$, for the $0$-$15$ mK range of continuum energies.  The data shows a good qualitative agreement with quantum chemistry calculations, and reveals a previously predicted \cite{KochGonzalezFerezPRA12_PAShapeResonanceControl} but so far unobserved $g$-wave shape resonance (or quasibound state) confined by the $J=4$ centrifugal barrier.  This long-lived ($\sim10$ ns) resonance 66(3) MHz above threshold could be used to control light-assisted molecule formation rates \cite{KochGonzalezFerezPRA12_PAShapeResonanceControl}.  Shape resonances can also be mapped with magnetic Feshbach dissociation of ground state molecules \cite{RempeVolzPRA05_ShapeResFeshbachSpectroscopy,GrimmKnoopPRL08_MetastableFeshbachMolecules}.  However, the photodissociation technique is widely applicable to molecules with any type of spin structure in any electronic state, and allows complete control over all quantum numbers.  In Fig. \ref{fig:Fig5}(b), an anisotropic, energy independent pattern is visible on all images, with a radius close to that of the 62 MHz image.  This signal arises from spontaneous photodissociation of the molecules into the $g$-wave shape resonance, and its angular pattern depends on $M_i$ \cite{Supplemental}.

This work explores light-induced molecular fragmentation in the fully quantum regime.  Quasiclassical descriptions are not applicable, and the observations are dominated by coherent superpositions of matter waves originating from monoenergetic fragments with different quantum numbers.  The results agree with a state-of-the-art quantum chemistry model \cite{MoszynskiSkomorowskiJCP12_Sr2Dynamics}, but challenge the theory to describe more complicated phenomena. For example, preliminary observations of quantum-state-resolved photodissociation to the doubly excited $^3P_1+{^3P}_1$ continuum (as in Fig. \ref{fig:Fig3}(b)) indicate rich structure near the threshold.  This continuum is not well understood, while interactions near the $^3P_1+{^3P}_1$ threshold play a key role in recent proposals and experiments in ultracold many-body science \cite{YeZhangScience14_SrSUNSymmetry}.  Photodissociation can shed light on the ultracold chemistry of a rich array of molecular states, as well as on new reaction mechanisms, as was shown here with M1/E2 photodissociation.  With an improved control of imaging and of the optical lattice effects, the experiments can get even closer to the threshold.  We expect to reach nK fragment energies in the lattice, leading to high precision measurements of binding energies for tests of fundamental physics \cite{GrimmBartensteinPRL05_Li2BindingEnergies,UbachsSalumbidesPRD13_FifthForcesMolecPrecisMeasts}.  Ultralow fragment energies can also aid in the creation of novel ultracold atomic gases \cite{LanePRA15_HFromBaH}.  A promising future direction is to enhance the quantum control achieved here by manipulating the final continuum states with external fields \cite{DoyleLemeshkoMP13_MoleculeManipulationEMFields,SeidemanStapelfeldtRMP03_AligningMoleculesStrongLaserPulses}.  We have shown an extreme sensitivity of weakly bound molecules to small magnetic fields \cite{ZelevinskyMcGuyerPRL15_Sr2ForbiddenE1}, and the same principle applies just above the threshold.  This external control over the ultracold chemistry should allow us to access and manipulate new reaction pathways.  Finally, we note that ultracold, quantum-state-resolved molecular fragmentation can serve as a probe for precise molecular spectroscopy \cite{GrimmMarkPRL07_StuckelbergInterferometryUltracoldCs2} or as a source of entangled states and coherent matter waves for a wide range of experiments in quantum optics \cite{AspectGrangierPRL85_Ca2PhotodissInterference,DrummondPRL05_EPRCorrelationsBECDissociation}.

\section{Acknowledgments}

We gratefully acknowledge ONR grant N000-14-14-1-0802, NIST award 60NANB13D163, and NSF grant PHY-1349725 for partial support of this work, and thank A. T. Grier, G. Z. Iwata, and M. G. Tarallo for helpful discussions.  R. M. acknowledges the Foundation for Polish Science for support through the MISTRZ program.

\onecolumngrid
\appendix*

\section{Supplementary information}

\subsection{Photodissociation of $0_\text{u}^+$ molecules}

\begin{figure}
	\centering
	\includegraphics[width=0.62\textwidth]{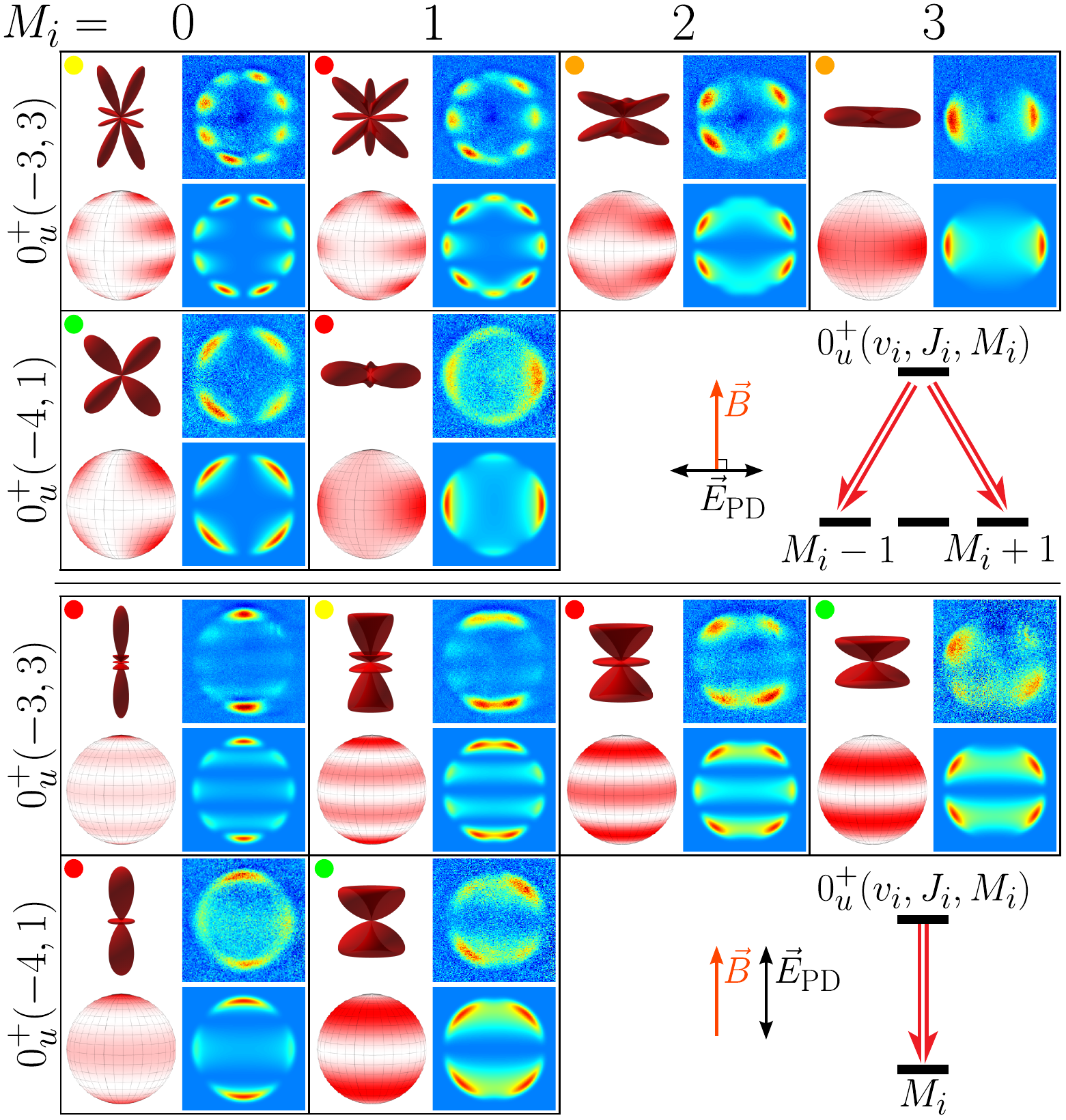}
	\caption{Additional absorption images and simulations of $^{88}$Sr$_2$ near the $^1S + {^3P_1}$ threshold fragmenting to the ground state continuum.  Here, the initial states are $0_\text{u}^+$ with $(v_i,J_i)$ = $(-4,1)$ or $(-3,3)$ as indicated.
Compatibility with the quasiclassical approximation is indicated by the colored dots.
}
	\label{fig:0u}
\end{figure}
In addition to the $(1)1_\text{u}$ states displayed in Fig.~4, we also measured the E1 photodissociation of $(1)0_\text{u}^+$ excited states to the ground continuum, as shown in Fig.~\ref{fig:0u}.

Note that in Figs. 4, \ref{fig:0u}, and \ref{fig:SE} the sign of $M_i$ does not affect the results, and our experiments used $M_i$ or $-M_i$ for some of the data sets.  To avoid confusion we did not label the figures with $|M_i|$, which suggests a superposition of $M_i$, but instead chose $M_i$ to be positive in the figures.

\subsection{Quasiclassical approximation}
\label{sec:Quasiclassical}

In the photodissociation literature there is a well-known ``quasiclassical'' approximation \cite{ZareBeswickJCP08_PhotofragmentAngularDistrQuantClass},
\begin{align}	\label{QC1}
I(\theta,\phi) \approx I_\text{qc}(\theta,\phi) = \Phi_i(\theta,\phi) \left[ 1 + \beta_{20} P_2^0(\cos \chi) \right],
\end{align}
that extends the conventional result (enclosed in brackets above) for the one-photon E1 photodissociation of a spherically symmetric initial state to other cases by multiplying with a probability density for the internuclear axis orientation of the initial bound state,
$\Phi_i(\theta,\phi)$.
Here, $\chi$ is the polar angle defined with respect to the orientation of linear polarization of the photodissociation light, while $(\theta,\phi)$ are fixed in the laboratory frame.

For homonuclear diatomic molecules in the Born-Oppenheimer approximation, the probability $\Phi_i(\theta,\phi)$ for an initial state with quantum numbers $J_i$, $M_i$, and $|\Omega_i|$ is given by Wigner-D functions as
\begin{align}	\label{QC2}
\Phi_i(\theta,\phi) =
	\frac{(2J_i+1)}{8\pi} \left( \left|D^{J_i}_{M_i \Omega_i}(0,\theta,0)\right|^2 + \left|D^{J_i}_{M_i,-\Omega_i}(0,\theta,0)\right|^2 \right),
\end{align}
where $\Omega$ is the internuclear projection of the electronic angular momentum, and the polar angle $\theta$ and azimuthal angle $\phi$ are defined by the quantization axis for $J_i$ and $M_i$, which may not be aligned with the linear polarization orientation of the photodissociation light.

We observe disagreement with the quasiclassical approximation in the majority of cases.
At first glance, this is surprising because theoretically, the quasiclassical approximation has been shown to be either equivalent or a good approximation to the quantum mechanical result for most cases of one-photon E1 photodissociation of a diatomic molecule with prompt axial recoil \cite{ZareBeswickJCP08_PhotofragmentAngularDistrQuantClass}.
However, our measurements are performed at very low continuum energies in order to reach the ultracold chemistry regime, and thus may violate the assumption of prompt axial recoil \cite{ZareMPC72_PhotoejectionDynamics}.
For example, the parameter $\delta$ in the interference amplitude of Fig.~5 is nonzero because the phase factor $(i)^{J} e^{i \delta_J}$ in Eq.~(\ref{t-definition}) depends on $J$, which the axial recoil approximation forbids.
Additionally, Ref.~\cite{ZareBeswickJCP08_PhotofragmentAngularDistrQuantClass} predicted that the quasiclassical approximation should fail for the special case of ``perpendicular'' transitions ($|\Delta \Omega| = 1$) with initial states that are a superposition of $\Omega_i$ states differing by $\pm 2$.
This special case includes our measurements of $1_\text{u}$ molecule photodissociation, and our observations support this prediction.

\begin{figure}
	\centering
	\includegraphics[width=0.58\textwidth]{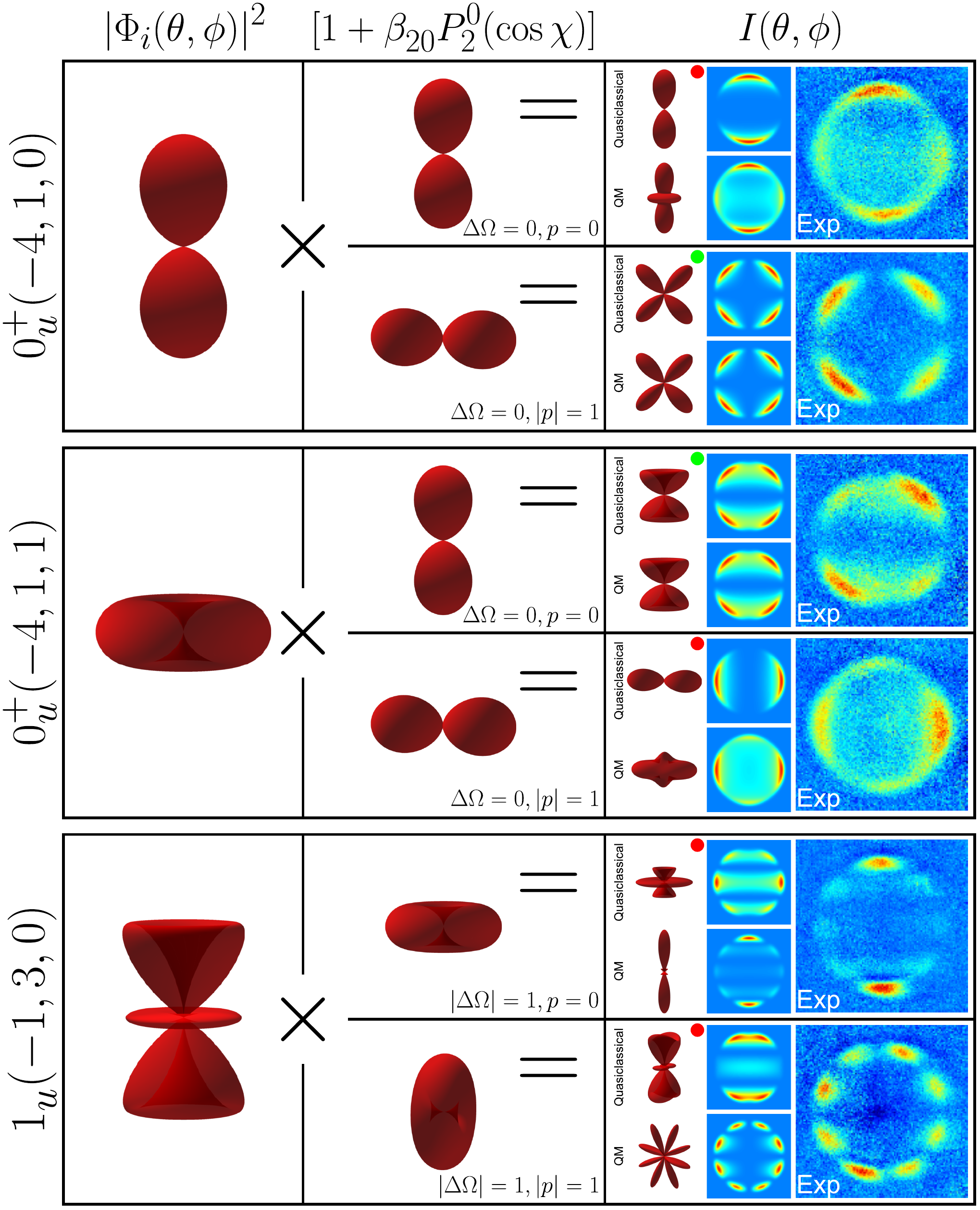}
	\caption{Comparison of quasiclassical and quantum mechanical (QM) theory with experimental (Exp) images for selected cases from Figs.~4 and \ref{fig:0u}.
The quasiclassical predictions follow from Eqs.~(\ref{QC1},\ref{QC2}) assuming $\beta_{20} = 2$ for $\Delta \Omega = 0$ and $-1$ for $|\Delta \Omega| = 1$.
(The quantum mechanical predictions slightly differ from those displayed in Figs.~4 and \ref{fig:0u} because they are the full quantum mechanical calculations given in Tables \ref{tabS2} and \ref{tabS3}.)
The colored dots indicate the level of quasiclassical agreement, as described in the text.
}
	\label{fig:QC}
\end{figure}
Figure \ref{fig:QC} compares the quasiclassical approximation with both quantum mechanical predictions and experimental images for several cases.
For each, the figure outlines the construction of the quasiclassical prediction.
As in Figs.~4 and \ref{fig:0u}, we use colored dots to indicate the level of agreement between the quasiclassical and quantum mechanical predictions, whose meaning is explained below.

In determining this agreement, we assumed that $|\Omega_i|$ was a good quantum number for the initial state while using the quasiclassical approximation, which is not exact for our excited-state molecules because of nonadiabatic Coriolis mixing \cite{ZelevinskyMcGuyerPRL13_Sr2ZeemanNonadiabatic}.
Additionally, note that there is some ambiguity in choosing a value of $\beta_{20}$ to use with the quasiclassical approximation.  Conventionally,  $\beta_{20}$ should be equal to $2$ for ``parallel'' transitions with $\Delta \Omega = 0$ and to $-1$ for ``perpendicular'' transitions with $|\Delta \Omega | = 1$.
As a first step we followed this conventional scheme, but if there was disagreement we next considered the effects of varying $\beta_{20}$ as a free parameter within the physically allowed range of $[-2,1]$.
Such a variation has been considered previously as an effect of the breakdown of the axial recoil approximation $\beta_{20}$ \cite{AshfoldWredeJCP02_AxialRecoilBreakdown}.
Similarly, there is ambiguity in determining the level of agreement for the cases that depend on the continuum energy (as in Figs.~2, 4, 5, \ref{fig:0u}).
Conventionally, the quasiclassical approximation assumes there is no such dependence because of the axial recoil approximation.
For Figs.~4, \ref{fig:0u}, and \ref{fig:QC} we chose to assign the colored dots by considering only the experimental images displayed, and ignoring other energies.

A green dot indicates an exact agreement with the quantum mechanical calculation.
We do observe three cases of exact agreement, which are all shown in Fig.~\ref{fig:0u}, two of which are highlighted in Fig.~\ref{fig:QC}.
The reason the quasiclassical approximation gives exact results is that selection rules only allow a single $J$ in these cases.
As a result, the axial recoil approximation is no longer necessary to derive the quasiclassical approximation.
Specifically, these cases correspond to $0_\text{u}^+$ initial states with odd $J_i$ for either $|M_i|=J_i$ with $p=0$ or $J_i=1$ and $M_i=0$ with $|p|=1$, for which the angular distribution is energy independent.
Agreement occurred without needing to adjust $\beta_{20}$ for these cases.

A yellow dot indicates qualitative agreement that cannot be made exact by adjusting the $\beta_{20}$ parameter in the quasiclassical expression, and
an orange dot indicates disagreement that can become a qualitative agreement by adjusting $\beta_{20}$.  A red dot indicates a clear disagreement for all values of $\beta_{20}$, usually because of incompatible extrema in $\Phi_i(\theta,\phi)$ and $I(\theta,\phi)$ (e.g., fragments observed where $\Phi_i(\theta,\phi)$ has a node).

In Fig.~2, the initial state $J_i=0$ is spherically symmetric and the angular distribution is parametrized only by $\beta_{20}$, so the quasiclassical approximation can always be adjusted to agree with the data at any continuum energy.

\subsection{Spontaneous photodissociation}

\begin{figure}
	\centering
	\includegraphics[width=1\textwidth]{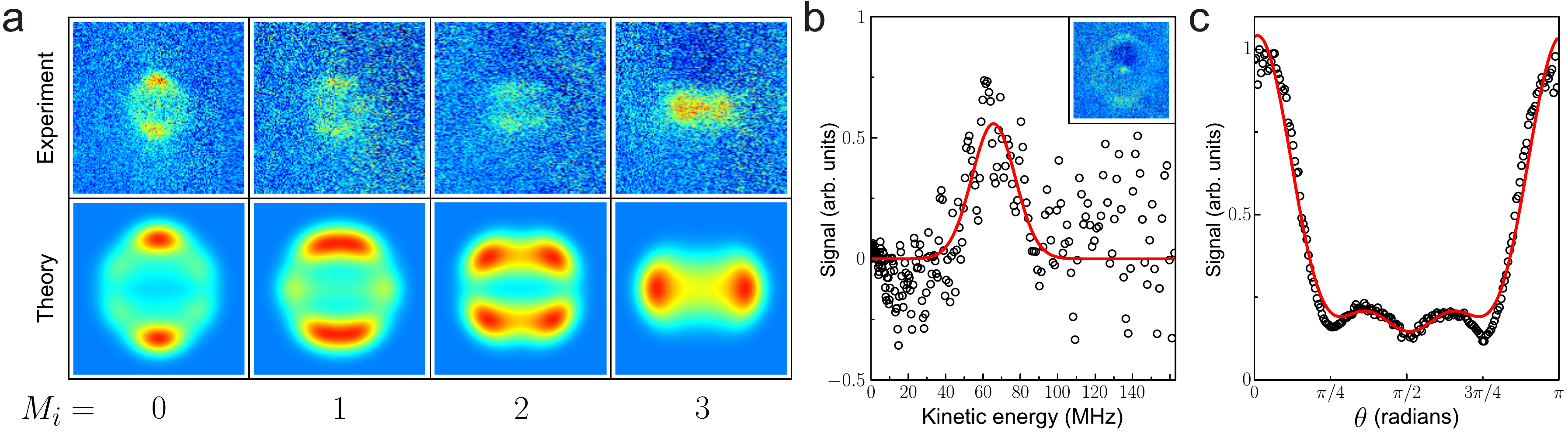}
	\caption{Spontaneous photodissociation of $^{88}$Sr$_2$ molecules, in the $0_\text{u}^+(v_i=-3,J_i=3,M_i)$ quantum state.  a) Absorption images of angular distributions versus $M_i$.  Theoretical simulations using Eqs.~(\ref{SEtheory}) and (\ref{MakeTheoryImages}) are shown underneath.  A short expansion time was used to increase visibility. 
b) For quantitative analysis, another image (inset) of the $M_i=0$ case was taken with a longer expansion time and analyzed with the pBasex algorithm.
The extracted radial distribution of photofragments shows a focusing of the fragments around a certain kinetic energy, which was determined by fitting with a Gaussian (red curve).
Correcting for an offset due to the lattice depth \cite{ApfelbeckThesis}, this energy corresponds to a shape resonance with a binding energy of $-$66(3) MHz.
c) The extracted angular distribution of photofragments matches the shape (red curve) expected from Eq.~(\ref{SEtheory}).
}
	\label{fig:SE}
\end{figure}
Figure \ref{fig:SE}(a) contains absorption images of the atomic fragments after spontaneous decay of the excited state $0_\text{u}^+(v_i=-3,J_i=3,M_i)$ to the ground continuum.
Since we selectively populate individual $M_i$ sublevels, the measured distributions are anisotropic.
They are well described by the incoherent superposition
\begin{align}	\label{SEtheory}
I(\theta) \approx \sum_{M} \left| Y_{4 M}(\theta,\phi) \right|^2
\begin{pmatrix}
4 & 1 & 3 \\
-M & M-M_i & M_i
\end{pmatrix}^2.
\end{align}
Here, $J$ is restricted to 4 because the strongest decay path is to an X($J=4$) shape resonance.
Note that if all $M_i$ were equally populated, which would add a sum over $M_i$ to Eq.~(\ref{SEtheory}), then the resulting distribution would be isotropic.

The shape resonance aids our measurement of the angular distributions because it favors a narrow range of continuum energies.  Figure \ref{fig:SE}(b) contains the results of pBasex analysis of the inset image, highlighting how the radial distribution of the atomic fragments agrees with expected energy and lifetime of the shape resonance.  Likewise, Fig.~\ref{fig:SE}(c) shows that the angular distribution from this analysis also matches expectations from Eq.~(\ref{SEtheory}).

\subsection{Parametrizing angular distributions}

\subsubsection{Beta parameters}

We can represent the angular distribution of any physical intensity (or differential cross section) with the expansion
\begin{align}	\label{exp1}
I(\theta,\phi) \propto 1 + \sum_{l=1}^\infty \sum_{m = 0}^l P_l^m(\cos \theta) \bigg[ \beta_{l m} \cos (m \phi) + \gamma_{l m} \sin (m \phi) \bigg]
\end{align}
in terms of real-valued ``anisotropy'' coefficients $\beta_{lm}$ and $\gamma_{lm}$, where $\gamma_{l0} \equiv 0$.
If there is no dependence on $\phi$ the associated Legendre polynomials reduce to Legendre polynomials, $P_l^0(\cos\theta) = P_l(\cos\theta)$, and the remaining coefficients $\beta_{l0}$ are conventionally denoted $\beta_l$.

If both atomic fragments of a dissociated homonuclear diatomic molecule are detected equally, then conservation of momentum requires the inversion symmetry 
$I(\pi - \theta, \phi + \pi) = I(\theta,\phi).$
The expansion (\ref{exp1}) has this symmetry if the coefficients with odd $l$ are zero.

If the only nonzero coefficients are those with even $l+m$, then the expansion (\ref{exp1}) will additionally be symmetric under reflection across the equator,
$I(\pi -\theta,\phi) = I(\theta,\phi).$
For homonuclear diatomic molecules, this symmetry requires the coefficients with odd $m$ to be zero.
Intensities without this symmetry display a ``skewness,'' such as the asymmetry in Fig.~1(b) and several other figure insets that are likely due to imperfect laser polarization.

\subsubsection{Partial scattering amplitudes}

For our experiments, the measured intensity (or differential cross section) can be written as a sum of the squared absolute values of complex scattering amplitudes for separate electronic channels,
\begin{align}	\label{Iff}
I(\theta,\phi) = \sum_{\Omega} \left| f^{\{\Omega\}} (\theta,\phi) \right|^2,
\end{align}
here indexed by the quantum number $\Omega$ for the internuclear projection of angular momentum.

To calculate the intensity, we compute the partial scattering amplitudes $f_{JM}^{\{\Omega\}}$ of a partial-wave expansion of the scattering amplitude in terms of angular basis functions,
\begin{align}	\label{PartialWavesExp}
f^{\{\Omega\}}(\theta,\phi) =  \sum_{JM} f_{J M}^{\{\Omega\}} \, \psi_{J M}^{\{\Omega\}}(\theta,\phi),
\end{align}
as described in Sec.~\ref{sec:PDTheory}.
In terms of Wigner D-functions, we chose the angular basis functions to be
\begin{align}
\psi_{J M}^{\{\Omega\}}(\theta,\phi) = \sqrt{\frac{2 J + 1}{4 \pi}} \, D^{J*}_{M \Omega}(\phi,\theta,0),
\end{align}
so that for $\Omega = 0$ they are equivalent to spherical harmonics, $\psi_{J M}^{\{0\}}(\theta,\phi) = Y_{J M}(\theta,\phi)$.

The expansion (\ref{exp1}) is equivalent to Eq.~(\ref{Iff}) if we write the anisotropy coefficients as the real and imaginary parts of a weighted sum over products of pairs of partial scattering amplitudes,
\begin{align}	\label{BGdensity}
\beta_{lm} + i \gamma_{lm} = \sum_{\Omega,J,J',M} W_{l m}(\Omega,J, J', M)
	\, \,   f_{J M}^{\{\Omega\}*} \, f_{J', M - m}^{\{\Omega\}}  \bigg/ \sum_{J M \Omega}  \left|f_{J M}^{\{\Omega\}} \right|^2.
\end{align}
The real-valued weights may be written in terms of Wigner 3j symbols as
\begin{align}	\label{Wlm}
W_{l m}(\Omega,J,J',M) = (-1)^{M-\Omega}
\frac{2 [l]}{1 + \delta_{m 0}}
\sqrt{\frac{ [J][J'](l-m)!}{(l+m)!}}
\begin{pmatrix}
J' & l & J \\
m-M & -m & M
\end{pmatrix}
\begin{pmatrix}
J' & l & J\\
-\Omega & 0 & \Omega
\end{pmatrix},
\end{align}
where the shorthand $[J]\equiv2J+1$ and $\delta_{i j}$ is a Kronecker delta.
As an aside, note that the quantities $ \rho_{M,M-m}^{\{\Omega J J' \}} = f_{J M}^{\{\Omega\}*} \, f_{J', M - m}^{\{\Omega\}}  \big/ \sum_{J M \Omega}  \left|f_{J M}^{\{\Omega\}} \right|^2$ in Eq.~(\ref{BGdensity}) have properties similar to density matrix elements.

From these weights $W_{lm}$ and the properties of 3j symbols, the maximum value of $l$ contributing in the expansion (\ref{exp1}) is limited to twice the largest value of $J$ for which there is a nonzero $f_{JM}^{\{\Omega\}}$.
The maximum value of $m$, is limited by the furthest off-diagonal magnetic coherence, that is, the nonzero quantity
$f_{J M}^{\{\Omega\}*} \, f_{J', M - m}^{\{\Omega\}}$ with largest $m = M-M'$.

\subsection{Experimental conditions and notes}
\label{ExptSection}

In all measurements, the photodissociation laser propagates along the tight-confinement $x$ axis of the optical lattice, and is linearly polarized along either the $y$ axis or the $z$ axis. Except for Fig.~2, for which the total magnetic fields is nearly zero, a magnetic field of a few to a few tens of Gauss is applied along the $z$ axis in order to define a quantization axis for excited bound and continuum states.  The ground bound and continuum states are only weakly sensitive to this field, so to avoid mixed-quantization effects from tensor light shifts \cite{ZelevinskyMcGuyerPRL15_Sr2ForbiddenE1} the optical lattice was linearly polarized along the $z$ axis.

After molecules are prepared in the quantum state of interest, the photodissociation transition is driven by a $\sim20$ $\mu$s laser pulse.
For electric dipole (E1) transitions under these conditions, the lab-frame spherical tensor components of the field driving the transition are
\begin{align}
T^1_0({\bf E}) &= E_z \\
T^1_{\pm1}({\bf E}) &= - \frac{i}{\sqrt{2}} E_y,
\end{align}
using the notation of Ref.~\cite{BrownCarrington}.
For linear polarization parallel to the $z$ axis, which is labeled ``$p=0$'' in Fig.~4, $E_y =0$.
For linear polarization along the $y$ axis, which is labeled ``$|p|=1$'' in Fig.~4, $E_z=0$.
Likewise, for magnetic dipole (M1) transitions these components are
\begin{align}
T^1_0({\bf B}) &= B_z \\
T^1_{\pm1}({\bf B}) &= - \frac{i}{\sqrt{2}} B_y.
\end{align}
Note that in Fig.~3 ``$p=0$'' now corresponds to linear polarization along the $y$ axis, such that $B_y=0$, and ``$|p|=1$'' to linear polarization parallel to the $z$ axis, such that $B_z=0$.
For electric quadrupole (E2) transitions, these components are
\begin{align}
T^2_0({\bf \nabla E}) &= 0 \\
T^2_{\pm1}({\bf \nabla E}) &= \mp \frac{1}{2} (i k E_z) \\
T^2_{\pm2}({\bf \nabla E}) &= \pm \frac{i}{2} (i k E_y)
\end{align}
for traveling-wave light propagating along the $x$ axis with wavenumber $k$ \cite{AuzinshBudkerRochester}.
For $\theta$ and $\phi$ defined as in Fig.~1, these experimental conditions produce angular distributions that can be described only with $\beta_{lm}$ coefficients in the expansion (\ref{exp1}).

After the photodissociation laser pulse, the photofragments are allowed to expand kinetically for several hundred $\mu$s before their positions are recorded with standard absorption imaging \cite{GueryOdelinReinaudiOL07_AbsImagingDenseAtomClouds}.  This expansion time is needed to mitigate blurring due to the finite pulse width and finite imaging resolution, but has the cost of diluting the signal over a larger area, which makes imaging artifacts more significant.

Most absorption images were taken with imaging light aligned very nearly along the $x$ axis.  These axial images are 2D projections of the 3D distribution of photofragment positions in the $yz$ plane.
To improve the signal to noise ratio, several hundred absorption images were averaged to produce a final record of the photofragment positions.
To remove imaging artifacts and incidental absorption from unwanted atoms, the experimental sequence was alternated so that every other image contained none of the desired atomic photofragments, but everything else.
The final image was then computed as the averaged difference between these interlaced ``with atoms'' and ``without atoms'' images.

For angular distributions that are cylindrically symmetric (depend only on $\theta$), the polar basis set expansion (pBasex) algorithm \cite{GarciaRSI04_2DImageInversion} can extract the 3D distribution from 2D projections like our axial images.
We used the software implementation of the pBasex algorithm in Ref.~\cite{AvaldiOKeeffeRSI11_PhotoelectronVelocityMapImaging} to do this for Figs.~2, 5, and \ref{fig:SE}.
For low signal-to-noise images, we found that the extracted distribution is artificially skewed towards spherical symmetry \cite{ApfelbeckThesis}.  To eliminate this systematic error, we performed pBasex inversion on a background image made from the set of ``without atoms'' images that is processed to remove imaging artifacts and rescaled so that the average bit depth equals that of the background regions in the final image.  The final distribution is then the difference between those extracted for the final image and for the background image.
The parameters $\beta_{20}$ of Fig.~2 and $R$ and $\delta$ of Fig.~5 were determined from least squares fitting of the number of photofragments versus $\theta$ in the final distribution.

For Fig.~2, additional analysis was performed by integrating 2D projections along $y$ to convert the images to 1D curves along $z$.
This allows parameters like $\beta_{20}$ to be directly extracted by fitting the 1D curve with the expected angular distribution, as in Fig.~\ref{fig:SE}(c).
While this analysis can be performed with the axial images, for Fig.~2 we did this through separate experiments with images taken with a camera facing along the $y$ axis, which had the benefit of a reduced optical depth.
These side-view images are 2D projections of the photofragment position onto the $xz$ plane, and are complicated by the distribution of occupied sites in the optical lattice.

\subsection{Theoretical calculation of angular distributions}

\subsubsection{Theoretical description of photodissociation}
\label{sec:PDTheory}

The theory of photodissociation employed in the manuscript follows the seminal work of Ref.~\cite{ZareMPC72_PhotoejectionDynamics}.
The fragmentation process is characterized by the differential cross section that is defined by Fermi's golden rule with the electric dipole (E1), magnetic dipole (M1), or electric quadrupole (E2) transition operators.
Since we work with a coupled manifold of electronic states for both the ungerade bound states and ungerade continuum, we do not assume the Born-Oppenheimer approximation in contrast to Zare \cite{ZareMPC72_PhotoejectionDynamics}.
In this case the theory of photodissociation for diatomic molecules is very similar to the non-degenerate atom-diatom case treated in detail by Balint-Kurti and Shapiro \cite{ShapiroBalintKurtiCP81_TriatomicPhotofragmentation,ShapiroBalintKurti_QuantTheoryPhotodissociation}.

In the absence of external fields, the wave function of the initial (bound) state depends on the set of the electronic coordinates $\{\textbf{r}\}$ and on the vector $\textbf{R}=(R,\Theta,\Phi)$ describing the relative motion of the nuclei, and is given by
\begin{align}	\label{Psi-initial}
\Psi ^{p_i}_{J_iM_i} (\{\textbf{r}\},\textbf{R})   &=  \nonumber \\
\sum_{\Omega_i=-J_i}^{J_i} &
\sqrt{\frac{2J_i+1}{16 \pi^2 (1+\delta_{\Omega_i 0})}} \left( D^{J_i \star}_{M_i \Omega_i} (\Phi,\Theta,0) \psi^{p_i}_{J_i \Omega _i} (\{\textbf{r}\},R) + \sigma_i D^{J_i \star}_{M_i, -\Omega_i} (\Phi,\Theta,0) \psi^{p_i}_{J_i, -\Omega _i} (\{\textbf{r}\},R) \right),
\end{align}
where $\sigma_i = p_i (-1)^{J_i}$ is the spectroscopic parity and $J_i$, $M_i$, and $p_i$ are the quantum numbers of the total angular momentum, its projection on the space-fixed $Z$ axis (previously denoted $z$), and the parity with respect to space-fixed inversion.
Note that in the above expression the quantum number $\pi_i$ related to the action of the reflection in the body-fixed $yz$ plane on the electronic coordinates does not appear.
In our case it is equal to zero, and the parity of the $\Omega_i=0$ electronic states is always ``+''.

In Hund's case (c) the internal wave function $\psi^{p_i}_{J_i \Omega _i} (\{\textbf{r}\},R)$ can be
represented by the Born-Huang expansion \cite{BornHuang,MoszynskiBusseryHonvaultJCP06_Ca2AbInitio}
\begin{align}	\label{psi-internal}
\psi^{p_i}_{J_i \Omega _i} (\{\textbf{r}\},R)=\sum_{n}\phi_{n,\Omega_i}(\{\textbf{r}\};R)\chi_{nJ_i \Omega _i}^{p_i}(R).
\end{align}
Here, the $\phi_{n,\Omega_i}(\{\textbf{r}\};R)$ are electronic wave functions, that is, the solutions of the electronic Schr\"odinger equation including spin-orbit coupling, which depend parametrically on the interatomic distance $R$.
The $\chi_{nJ_i \Omega _i}^{p_i}(R)$ are rovibrational wave functions.
Finally, the index $n$ labels all relativistic dissociation channels.
Note that for homonuclear diatomic molecules the electronic wave function has an additional gerade/ungerade (g/u) symmetry resulting from the $D_{\infty h}$ point group of the molecule.
For simplicity we do not indicate the g/u symmetry in the notation $\phi_{n,\Omega_i}(\{\textbf{r}\};R)$.
The rovibrational wave functions are solution of a system of coupled differential equations.
See, for instance, Ref.~\cite{MoszynskiSkomorowskiJCP12_Sr2Dynamics} for the equations corresponding to the ungerade excited manifold of the electronic states.

The wave function $\Psi_\textbf{k}^{p_f}(\{\textbf{r}\}, \textbf{R})$ of the final continuum state corresponding to the wave vector $\textbf{k}=(k,\theta,\phi)$ can be represented by the following expansion reflecting different partial waves $J$ of the fragmented atoms,
\begin{align}	\label{Psi-final}
\Psi_\textbf{k}^{p_f}(\{\textbf{r}\},\textbf{R})  =&
\sum_{J M}
\sum_{\Omega =-J}^{J} \sum_{\Omega' =-J}^{J}
\frac{(2 J +1)}{4 \pi \sqrt{2 \pi (1+\delta_{\Omega' 0})}}
	\; \mathcal{C}^{j p}_{m_j \Omega} D^{J}_{M \Omega} (\phi,\theta,0) D^{J*}_{m_J \Omega} (\phi,\theta,0) \nonumber
\\
	& \times \left( D^{J \star}_{M \Omega'} (\Phi,\Theta,0) \psi^{J p_f}_{\Omega' \Omega} (\{\textbf{r}\},R) + \sigma_i D^{J \star}_{M, -\Omega'} (\Phi,\Theta,0) \psi^{J p_f}_{-\Omega', \Omega } (\{\textbf{r}\},R) \right),
\end{align}
where $j$ denotes the total angular momentum of the photofragmented atoms, $m_j$ is its projection in the space-fixed $Z$ axis, and $p$ is the product of atomic parities.
The numerical coefficients $\mathcal{C}_{m_j \Omega}^{j p}$ depend on the states of the photofragmented atoms and can be found in Ref.~\cite{VasyutinskiiKupriyanovCP93_PhotofragmentOrientationAlignment}.
The internal wave function is given by the following multichannel generalization of the Born-Huang expansion,
\begin{align}	\label{psi-mc}
\psi^{J p_i}_{\Omega' \Omega} (\{\textbf{r}\},R)=\sum_{n} \phi_{n,\Omega'}(\{\textbf{r}\};R) \chi_{n \Omega' \Omega}^{J p_f}(R),
\end{align}
where $\chi_{n \Omega' \Omega}^{J p_f}(R)$ is a radial channel function that satisfies the boundary condition
\begin{align}	\label{bound-mc}
\chi_{n \Omega' \Omega}^{J p_f}(R)\approx R \sqrt{\frac{2 \mu k }{\pi }} \left( \delta_{\Omega' \Omega} j_J(k R)  +  S_{n \Omega' \Omega}^{ J p_f} n_J(k R) \right)\quad (\text{for}~R \longrightarrow \infty)
\end{align}
in terms of the scattering matrix $S _{n \Omega' \Omega}^{ J p_f}$.  Here, $\mu$ is the reduced mass and $j_J$, $n_J$ are spherical Bessel functions.

In this work we considered four different photofragmentation processes:
(i) the E1 process starting from the ungerade manifold of the electronic states that correspond to the $^1S+{^3P}_1$ dissociation limit and ending at the ground electronic continuum,
(ii) the M1 and (iii) the E2 processes starting from the gerade manifold corresponding to the same dissociation limit and ending at the ground electronic continuum,
and finally (iv) the E1 process starting from ground state molecules and ending at the ungerade manifold corresponding to the $^1S+{^3P}_1$ dissociation limit.

The first three processes begin with manifolds that are described by two coupled electronic states: $0_\text{u}^+$ and $1_\text{u}$ for the E1 process and $0_\text{g}^+$ and $1_\text{g}$ for the M1 and E2 processes.
The corresponding wave functions for the initial states are given by Eqs.~(\ref{Psi-initial}) and (\ref{psi-internal}) with the summation over $\Omega_i$ limited to $-1$, 0, and 1, and with $n$ fixed to $^1S+{^3P}_1$.
The wave function for the final continuum state, however, corresponds to the Born-Oppenheimer approximation and is given by Eqs.~(\ref{Psi-final}) and (\ref{psi-internal}) with only the $\Omega = 0$ term and with $n$ fixed to $^1S+{^1S}$.
Note that in the single-channel approximation $\Omega = \Omega'$, so for simplicity we denote the rovibrational wavefunction by $\chi_{n \Omega}^{J p_f}(R)$.
The electronic transition operator for the E1 process was assumed to be constant and proportional to the atomic value, while the operators for the M1 and E2 transitions followed the asymptotic form of Refs.~\cite{MoszynskiBusseryHonvaultMP06_Ca2AbInitio,ZelevinskyMcGuyerNPhys15_Sr2M1}.
Otherwise, the remaining derivation of the expression for the differential cross section follows Ref. \cite{ZareMPC72_PhotoejectionDynamics} and is not reproduced here, although the multichannel character of the initial state wave functions complicates the angular momentum algebra.

Now we discuss the boundary condition for the final continuum rovibrational wave function $\chi_{n \Omega }^{J p_f}(R)$.
The single-channel approximation is valid for the $0_\text{g}^+$ ground electronic continuum.
In this case, at large internuclear distances $R$ the partial wave expansion (\ref{Psi-final}) becomes \cite{Levine}
\begin{equation}	\label{chi-bound}
\chi_{n \Omega}^{J p_f}(R)\approx i^{J} e^{i \delta_{J}} \sqrt{\frac{2 \mu }{\pi \hbar^2 k}} \sin(kR+\delta_{J}+J \pi)\quad (\text{for}~R \longrightarrow \infty),
\end{equation}
where $\delta_{J}$ is the phase shift for a given partial wave $J$.
However, in practice it is more convenient to work with real functions than with complex functions that satisfy this boundary condition.
Therefore, we chose to instead use the real-valued large-$R$ boundary condition \cite{Levine}
\begin{align}
\chi_{n \Omega}^{J p_f}(R)\approx \sqrt{\frac{2 \mu }{\pi \hbar^2 k}} \sin(kR+\delta_{J}+J \pi)\quad (\text{for}~R \longrightarrow \infty),
\end{align}
and to include the phase factor $ i^{J} e^{i\delta_{J}}$ in the partial scattering amplitudes.

Therefore, for the first three processes the differential cross section is given by Eq.~(\ref{Iff}) with $\Omega = 0$. The partial scattering amplitudes in the expansion (\ref{PartialWavesExp}) for this scattering amplitude are then given by
\begin{align} 	 \label{t-definition}
f_{J M}^{\{\Omega\}} &= \sum_{\Omega_i=-J_i}^{J_i}
2 i^{J} e^{i\delta_{J}}
\sqrt{\frac{4 \pi (2J_i+1)}{(1+\delta_{\Omega_i 0})(1+\delta_{\Omega 0})}}
\langle \chi_{n k J \Omega}^{p_f} | T^L_{\Omega - \Omega_i} |\chi_{n J_i \Omega _i}^{p_i} \rangle
\begin{pmatrix}
J & L & J_i\\
-\Omega &\Omega-\Omega_i& \Omega_i
\end{pmatrix}
\begin{pmatrix}
J & L & J_i\\
-M &M-M_i& M_i
\end{pmatrix},
\end{align}
where $T^L_{\Omega - \Omega_i}$ is the electronic transition operator of rank $L=1$ for E1 or M1 transitions and $L=2$ for E2 transitions for the experimental conditions described in Sec.~\ref{ExptSection}.
The anisotropy parameters in the expansion (\ref{exp1}) then follow from using these partial amplitudes with Eqs.~(\ref{BGdensity}) and (\ref{Wlm}).

The M1 and E2 processes were not observed separately because of selection rules.
In this case, both processes must be included and the observed cross section may be written as the sum
\begin{align}	 \label{exp-amplitude}
I(\theta,\phi) = \left| f_{J M, \text{M1}}^{\{0\}} + f_{J M, \text{E2}}^{\{0\}} \right|^2,
\end{align}
of separate scattering amplitudes (\ref{PartialWavesExp}) using Eq.~(\ref{t-definition}).
This expression explicitly allows for interference between the M1 and E2 processes.  Note that this interference may affect the angular distribution even if it does not affect the strength of the transition, which is proportional to the integral of the differential cross section over all angles.

Finally, for the fourth process of photofragmentation beginning with ground state molecules and ending at the $^1S +{^3P}_1$ continuum, the wave function of the initial (bound) state satisfies the Born-Oppenheimer approximation.
Therefore we set $\Omega_i = 0$ in Eq.~(\ref{Psi-initial}) and limit the Born-Huang expansion (\ref{psi-internal}) to a single product.
However, the partial wave expansion (\ref{Psi-final}) for the final continuum must explicitly account for the Coriolis coupling between the $0_\text{u}^+$ and $1_\text{u}$ electronic states, and for the angular momentum $j=1$ and total parity $p=-1$ of the atomic fragments.
For this multichannel continuum case, we imposed the complex boundary conditions of Eq.~(\ref{bound-mc}).
The remaining derivation of the expression for the differential cross section follows Refs.~\cite{ZareBeswickJCP08_PhotofragmentAngularDistrQuantClass,ShapiroBalintKurtiCP81_TriatomicPhotofragmentation,ShapiroBalintKurti_QuantTheoryPhotodissociation,VasyutinskiiShterninJCP08_PhotofragmentsCoriolis}.
The differential cross section $I(\theta,\phi)$ then follows from Eqs.~(\ref{Iff}) and (\ref{PartialWavesExp}) using the partial scattering amplitudes
\begin{align} 	\label{t-definition-mc}
f_{J M}^{\{\Omega\}} 	
	&= \sum_{\Omega_i=-J_i}^{J_i} \sum_{\Omega'=-J_i}^{J_i}
2
\sqrt{\frac{4 \pi (2J_i+1)}{(1+\delta_{\Omega_i 0})(1+\delta_{\Omega' 0})}}
\langle \chi_{n \Omega' \Omega }^{ J p_f} | T^L_{\Omega' - \Omega_i} |\chi_{n J_i \Omega _i}^{p_i} \rangle
\begin{pmatrix}
J & j & J_i\\
-\Omega' &\Omega'-\Omega_i& \Omega_i
\end{pmatrix}
\begin{pmatrix}
J & j & J_i\\
-M &M-M_i& M_i
\end{pmatrix}.
\end{align}

\subsubsection{Energy independent angular distributions}

For $^{88}$Sr$_2$ photodissociation, there are a few cases where the angular distributions are independent of the continuum energy.
For E1 photodissociation to the ground continuum (where symmetry restricts $J$ to even values) this occurs if $J_i$ is even because $\Delta J = 0, \pm 1$.
This also occurs for odd $J_i$ for the cases described in Sec.~\ref{sec:Quasiclassical}.
For M1/E2 photodissociation to the ground continuum from $J_i = 1$, $M_i = 0$, this occurs when the dissociation laser is linearly polarized along the $z$ axis, because of $\Delta M$ selection rules.
In these cases, the energy-dependent radial integrals in Eq.~(\ref{t-definition}) are common to all partial scattering amplitudes, so the angular distributions are independent of the continuum energy.  They are also relatively simple to calculate, because they reduce to evaluating geometrical factors.

\subsubsection{Single $J$ approximation for angular distributions}

In a similar fashion, Eqs.~(\ref{t-definition}) and (\ref{t-definition-mc}) can be used to explore what range of angular distributions may occur in an experiment by making simplifying assumptions about the radial integrals.
For example, for E1 photodissociation to the ground continuum we could approximate the radial matrix elements in Eq.~(\ref{t-definition}) to be nonzero only for a single $J$ but multiple $M$, such that
\begin{align}	\label{Xapprox}
f_{J M}^{\{\Omega\}} \propto T^L_{M-M_i}({\bf E})
\begin{pmatrix}
J & L & J_i \\
-M & M-M_i & M_i
\end{pmatrix},
\end{align}
where $T^L_{M-M_i}({\bf E})$ are the lab-frame spherical tensor components of the dissociating field, as described in Sec.~\ref{ExptSection} for our experimental conditions.
For a selected initial state, the calculation of the angular distribution simplifies to evaluating geometrical factors that depend only on the allowed quantum numbers.

In addition to energy-independent cases, this approximation works well for the energy-dependent data in Fig.~4 with odd $J_i$, where the continuum energies were chosen so that a single $J$ was responsible for most of each angular distribution:  $J=4$ for $J_i=3,4$ and $J=2$ for $J_i=1,2$.
This approximation also explains some interesting properties that we observe.
For example, the $(M_i=0,|p|=1)$ cases are identical except for a $90^\circ$ rotation in $\phi$, which corresponds to alternating the sign of the $\beta_{l2}$ parameters with $J_i$.
For $M_i\neq0$, a qualitatively similar rotation often occurs.
Finally, for $|p|=1$ the cases of $|M_i| = J$ have the same coefficients as those of $p=0$ for $|M_i|= J - 1$, since they produce the same single sublevels $M$ in the continuum.

\subsection{Parameters for theoretical images in figures}

\begin{table}
\caption{\label{tabS1}Parameters $\beta_{lm}$ of the theoretical images in Fig.~3 for the $p=0$ case of M1/E2 photodissociation of $1_\text{g}(v_i=-1,J_i=1,M_i=0)$.}
\begin{center}
\begin{tabular*}{\textwidth}{@{\extracolsep{\fill}} c |  r r r r r }
\toprule[1pt]
$\epsilon/h$ (MHz) &
	$\beta_{20}$	& $\beta_{22}$	& $\beta_{40}$	& $\beta_{42}$	& $\beta_{44}$	\\
\midrule[.5pt]
8 &     0.2475          & $-$0.3563     & 0.3077                & $-$0.03626    & 0.006409 \\
\bottomrule[1pt]
\end{tabular*}
\end{center}
\end{table}
\begin{table}
\caption{\label{tabS2}Parameters $\beta_{lm}$ of the theoretical images in Figs.~4 and \ref{fig:QC} for E1 photodissociation of $1_\text{u}(v_i=-1,J_i,M_i)$.  
Values for odd $J_i$ that depend on the continuum energy $\epsilon$ (given below in MHz) are rounded to four significant figures.
However, the energies for odd $J_i$ were chosen so that Eq.~(\ref{Xapprox}) with $J=J_i+1$ is a good approximation.
Figure 4 uses the values from this approximation, which are given on a second line as fractions.  Figure \ref{fig:QC} uses the unapproximated values.
For even $J_i$, the values are independent of the continuum energy.
Where the transition is forbidden by symmetry ($p_i=M_i=0$, even $J_i$), the experimental pattern is enabled by an applied magnetic field that admixes excited states \cite{ZelevinskyMcGuyerPRL15_Sr2ForbiddenE1} such that the parameters correspond to the approximation (\ref{Xapprox}) with the substitution $J_i \longrightarrow J_i \pm 1$. 
Omitted values are zero because of symmetry.  }
\begin{center}
\scalebox{0.85}{
\begin{tabular*}{1.16\textwidth}{@{\extracolsep{\fill}} c c c  |  r r r r r r r r }
\toprule[1pt]
$J_i$ & $|M_i|$ & $\epsilon/h$&
	$\beta_{20}$	& $\beta_{22}$	& $\beta_{40}$	& $\beta_{42}$	& $\beta_{60}$	& $\beta_{62}$	& $\beta_{80}$	& $\beta_{82}$	\\
\midrule[.5pt]
\multicolumn{11}{c}{$|p|=1$}\\
\midrule[.5pt]
4 & 0 & 76		&
	85/77		& 25/77		& 729/1001		& 81/1001			&$-$1/11		&1/22 		& $-$392/143  	& 7/143 \\
4 & 1 & 74		&
	1360/1463	& 450/1463	& 6561/19019		& 81/1729			& 2/209		& 0			& 6664/2717 	&$-$105/2717 \\
4 & 2 & 71		&
	65/154		& 45/176		& $-$243/728		& $-$81/2288	 	& 5/8			& $-$9/176	& $-$245/143 	& 21/1144 \\
4 & 3 & 68		&
	$-$40/121		& 20/121		&$-$243/1573 		& $-$162/1573		& $-$170/121	& 4/121		& 1400/1573 	& $-$7/1573 \\
4 & 4 & ---		&
	$-$5/11		& 0  			& $-$243/143		& 0 				& 17/11		& 0 			& $-$56/143 	& 0 \\
3 & 0 & 71		&
	0.3097		& $-$0.3909	& 0.3584			& $-$0.05354		& 0.8934		& $-$0.009862	& $-$2.561 	& $-$0.04574 \\ 
&&&	
	85/77		& $-$25/77	& 729/1001		& $-$81/1001		&$-$1/11		&$-$1/22 		& $-$392/143  	& $-$7/143 \\
3 & 1 & 71	 	&
	0.1445		& $-$0.2850	&$-$0.1389		& $-$0.01507		&$-$1.653		& $-$0.007708	& 1.816		& 0.03244 \\
&&&	
	400/539		& $-$150/539	& $-$1539/7007	& $-$27/637		& $-$10/11	& 0			& 280/143 	& 5/143 \\
3 & 2 & 72	 	&
	$-$0.4966		&$-$0.1545	&$-$1.515			& 0.02140			& 1.633		& 0.03091		&$-$0.6213 	& $-$0.01110 \\
&&&	
	$-$20/77		& $-$15/88	& $-$5589/4004	& 27/1144			& 59/44 		& 3/88		& $-$98/143 	& $-$7/572 \\ 
3 & 3 & 72 	&
	$-$1.835		& $-$0.05559	& 1.208			& 0.03464			& $-$0.4569	& $-$0.01112	& 0.08381		& 0.001497 \\
&&&
	$-$3880/2233 & $-$20/319	& 30861/29029		& 162/4147		& $-$134/319	& $-$4/319	& 392/4147 	& 7/4147\\
2 & 0 & 52		&
	5/7		& 5/14		& $-$12/7 		& 1/7 \\
2 & 1 & 48		&
	2/7		& 2/7			& 12/7		&$-$3/35 \\
2 & 2 & 44		&
	5/7		& 0 			& $-$12/7 		& 0 \\
1 & 0 & 32		&
	5/7		& $-$5/14 		& $-$12/7 		& $-$1/7 \\ 
1 & 1 & 31		&
	$-$0.3445		& $-$0.1311	& 0.1921			& 0.01601 \\ 
&&&
	$-$50/49	& $-$10/49	& 36/49		& 3/49\\ 
\midrule[.5pt]
\multicolumn{11}{c}{$p=0$}\\
\midrule[.5pt]
4 & 0 & 77		&
	100/77 	&	& 1458/1001 	& 	& 20/11 	&	& 490/143 	&  \\
4 & 1 & 78		&
	85/77 	&	& 729/1001 	& 	& $-$1/11 	& 	& $-$392/143 	&   \\
4 & 2 & 71		&
	40/77 	&	& $-$81/91 	& 	& $-$2 	& 	& 196/143 	&   \\
4 & 3 & 68		&
	$-$5/11 	& 	& $-$243/143 	& 	& 17/11 	& 	& $-$56/143 	&   \\
4 & 4 & ---		&
	$-$20/11 	& 	& 162/143 	& 	& $-$4/11 	& 	& 7/143  		&  \\
2 & 0 & 56		&
	10/7		&	& 18/7		&  \\
2 & 1 & 55		&
	5/7		&	& $-$12/7 		&  \\
2 & 2 & 44		&
	$-$10/7 	& 	& 3/7 		& \\
\bottomrule[1pt]
\end{tabular*}}	
\end{center}
\end{table}
\begin{table}
\caption{\label{tabS3}Parameters $\beta_{lm}$ of the theoretical images in Figs.~\ref{fig:0u} and \ref{fig:QC} for E1 photodissociation of $0_\text{u}^+(v_i,J_i,M_i)$, with $v_i=-3$ for $J_i = 3$ and $v_i=-4$ for $J_i=1$.  
Only odd $J_i$ are allowed by symmetry.  The values that depend on the continuum energy $\epsilon$ (given below in MHz) are rounded to four significant figures.
However, as in Table \ref{tabS2}, the energies were chosen so that Eq.~(\ref{Xapprox}) with $J=J_i+1$ is a good approximation.
Figure \ref{fig:0u} uses these approximate values, which are given on a second line as fractions below.  
Figure \ref{fig:QC} uses the unapproximated values.
Omitted values are zero because of symmetry. }
\begin{center}
\scalebox{0.85}{
\begin{tabular*}{1.16\textwidth}{@{\extracolsep{\fill}} c c c  |  r r r r r r r r }
\toprule[1pt]
$J_i$ & $|M_i|$ & $\epsilon/h$&
	$\beta_{20}$	& $\beta_{22}$	& $\beta_{40}$	& $\beta_{42}$	& $\beta_{60}$	& $\beta_{62}$	& $\beta_{80}$	& $\beta_{82}$ \\
\midrule[.5pt]
\multicolumn{11}{c}{$|p|=1$}\\
\midrule[.5pt]
3 & 0 & 72 	&	
	0.5258		& $-$0.3728	& 0.4946		& $-$0.05980	& 0.6317		& $-$0.01999	& $-$2.652	& $-$0.04736 \\
&&&	
	85/77		& $-$25/77	& 729/1001	& $-$81/1001	&$-$1/11		&$-$1/22 		& $-$392/143  	& $-$7/143 \\
3 & 1 & 71 	&	
	0.3251		& $-$0.2842	& $-$0.2059	& $-$0.02529	& $-$1.437	& $-$0.005318	& 1.893		& 0.03381 \\
&&&	
	400/539		& $-$150/539	& $-$1539/7007 & $-$27/637	& $-$10/11	& 0			& 280/143 	& 5/143 \\
3 & 2 & 70 	&	
	$-$0.4253		& $-$0.1640	& $-$1.463	& 0.02271		& 1.548		& 0.03281		& $-$0.6596	& $-$0.01178 \\
&&&	
	$-$20/77		& $-$15/88	& $-$5589/4004 & 27/1144	& 59/44 		& 3/88		& $-$98/143 	& $-$7/572 \\ 
3 & 3 & 69 	&	
	$-$1.801		& $-$0.06003	& 1.161		& 0.03740		& $-$0.4509	& $-$0.01201	& 0.09050		& 0.001616 \\
&&&
	$-$3880/2233 & $-$20/319	& 30861/29029	& 162/4147	& $-$134/319	& $-$4/319	& 392/4147 	& 7/4147\\
1 & 0 & 33 	&	
	5/7			& $-$5/14		& $-$12/7 		& $-$1/7	\\
1 & 1 & 32 	&	
	$-$0.4751		& $-$0.03716 	& 0.3941		& 0.03284 \\
&&&
	$-$50/49		& $-$10/49	& 36/49		& 3/49 \\
\midrule[.5pt]
\multicolumn{11}{c}{$p=0$}\\
\midrule[.5pt]
3 & 0 & 72 	&	
	2.131		& 			& 2.272			& 				& 3.022		& 			& 3.223\\
&&&
	100/77		& 			& 1458/1001		& 				& 20/11		& 			& 490/143\\
3 & 1 & 73 	&	
	1.861		& 			& 0.7914			& 				& $-$1.079	& 			& $-$2.573\\
&&&
	85/77		& 			& 729/1001		& 				& $-$1/11		& 			& $-$392/143\\
3 & 2 & 74 	&	
	0.9400		& 			& $-$1.677		& 				& $-$1.561	& 			& 1.298\\
&&&
	40/77 		& 			& $-$81/91		& 				& $-$2		& 			& 196/143 \\
3 & 3 & 75 	&	
	 $-$5/11		& 			& $-$243/143		& 				& 17/11		& 			& $-$56/143\\
1 & 0 & 33 	&	
	0.4265		& 			& 1.035 \\
&&&
	 10/7			& 			& 18/7 \\
1 & 1 & 32 	&	
	5/7			& 			& $-$12/7\\
\bottomrule[1pt]
\end{tabular*}}	
\end{center}
\end{table}
Tables \ref{tabS1}, \ref{tabS2}, and \ref{tabS3} list the parameters used to generate the theoretical images shown in Figs.~3, 4, \ref{fig:0u},  and \ref{fig:QC}.
For reference, the binding energies for the $1_\text{u}(v_i=-1,J_i)$ excited states in MHz are 353 for $J'=1$, 287 for $J'=2$, 171 for $J'=3$, and 56 for $J'=4$;
for the $0_\text{u}^+(v_i=-4,J_i=1)$ state, 1084 MHz;
for the $0_\text{u}^+(v_i=-3,J_i=3)$ state, 132 MHz;
and for the $1_\text{g}(v_i=-1,J_i=1)$ state, 19 MHz.

To display theoretical results as simulated absorption images, the intensities are projected into the $yz$ plane by integrating over the $x$ direction.  To approximate the blurring present in experimental images from limited optical resolution and light pulse durations, the image is convolved with a Gaussian distribution,
\begin{align}	\label{MakeTheoryImages}
N(y,z) \propto \int_{-\infty}^\infty e^{-\left(\sqrt{x^2 + y^2 + z^2} - R_0 \right)^2/\left(2 \sigma^2 \right)}  \, I\left( \cos^{-1}\left[ z/\sqrt{x^2+y^2+z^2} \right],\sin^{-1} \left[ y/\sqrt{x^2+y^2} \right] \right) dx,
\end{align}
where $R_0$ is the mean radius and $\sigma$ is the the standard deviation.
For the theoretical images in the paper, the fractional blur was $\sigma/R_0 = 0.05$.


\end{document}